 \documentclass[authoryear,preprint,review,12pt]{elsarticle}

\usepackage{amssymb}

\journal{}

\begin{document}

\begin{frontmatter}



\title{Complex Broad Emission Line Profiles of AGN - Geometry of the Broad Line Region}


\author[1]{E. Bon}
\ead{ebon@aob.bg.ac.rs}

\author[1,2]{N. Gavrilovi\' c}
\ead{ngavrilovic@aob.bg.ac.rs}

\author[3]{G. La Mura}
\ead{giovanni.lamura@unipd.it}

\author[1]{L. \v C. Popovi\' c}
\ead{lpopovic@aob.bg.ac.rs}

\address[1]{Astronomical Observatory, Volgina7, 11060 Belgrade, Serbia}

\address[2]{Observatoire de Lyon, 9 avenue Charles Andr\'e, Saint-Genis
Laval cedex, F-69561, France ; CNRS, UMR 5574}

\address[3]{Department of Astronomy, University of Padova, Vicolo dell'Osservatorio, I-35122 Padova, Italy}

\begin{abstract}

The Broad Emission Lines (BELs) in spectra of type 1 Active Galactic Nuclei
(AGN) can be very complex, indicating a complex Broad Line Region (BLR) geometry. According to
the standard unification model  one can expect an accretion disk around a
supermassive black hole in all AGN. Therefore, a disk geometry is expected
in the BLR. However, a small fraction of BELs show double-peaked profiles
which indicate the disk geometry. Here, we discuss a two-component model,
assuming an emission from the accretion disk and one additional emission
from surrounding region. We compared the modeled BELs with observed ones
(mostly broad H$\alpha$ and H$\beta$ profiles) finding that the model can
well describe single-peaked and double-peaked observed broad line profiles.

\end{abstract}


\end{frontmatter}


\section{Introduction}

The detailed structure of the innermost part of AGN is still an open question. It is widely
accepted that the central engine consists of a super massive black hole fueled by an accretion disk. The origin
of the broad emission lines, observed in most AGN, is explained by photoionization of dense gas surrounding the nuclei,
but the structure and dynamics of the BLR remains unclear.
Using Broad Emission Lines (BEL) shape investigations of the BLR, one can constrain the geometry of the material in the BLR \citep[see][]{Pop06}.
 
BELs of AGN show high diversity in their shapes and widths. In some rare cases, where the accretion disk clearly contributes to the BELs, we can investigate the disk properties using a broad, double-peaked, low-ionization lines \citep{Che1,Che2,Era, Era1,Era09}. The rotation of the material in the disk results in one blue-shifted and one redshifted peak, while the gravitational redshift produces a displacement of the center of the line and a distortion of the line profile. 

Moreover, a number of BELs show shoulders-like profile in the wings of the line, which could be a fingerprint of the accretion disk emission \citep{Pop03,Era1, Str03}. Beside  emission of the  disk (or disk-like region) or emission from
spiral shock waves within a disk \citep{Che1,Che2},
the following geometries may cause substructures in line profiles:
i) emission from the oppositely-directed
sides of a bipolar outflow \citep{Zhen90,Zhen91};
ii) emission from a spherical system of clouds in
randomly inclined Keplerian orbits illuminated anisotropically from the
center \citep{GW96}; and
iii) emission of the binary black hole system \citep{Gas83,Gas96};

To explain the complex morphology of the observed BEL shapes, different geometrical models have been
discussed \citep[see in more details][]{Sul00}. In some cases the BEL profiles can be explained only
if two or more kinematically different emission regions are considered   \citep[see e.g.][]{Marziani93,Rom96,
Pop01, Pop02, Pop03, Pop04, Bon06, Ilic06, Col06, Hu08}. In particular, the existence of a Very Broad Line
Region (VLBR) with random velocities at ~5000-6000 km/s within an Intermediate Line Region (ILR) has
also been considered to explain the observed BEL profiles \citep{CB96,Sul00,Hu08}.

Nevertheless, the majority of AGN with BELs have only single peaked lines, but this does
not necessarily indicate that the contribution of the disk
emission to the BELs profiles is negligible.

In this paper we study the possibility that the flux of the hidden disk emission could be present
in single peaked  emission line profiles of AGN. With this goal,
we applied a simple, two-component model, which assumes that observed BELs can be represented by composition of the emission of the accretion disk and surrounding non-disk isotropic region. For the accretion disk emission, we have used the model proposed by \citet{Che1}, which assumes optically thick and geometrically thin disk, while for the additional emitting region which surrounds the disk, we assumed isotropic velocity distribution of emitting clouds, which imply that the emission line profiles generated by this region can be described by a Gaussian function.
In Section 2 of this paper we describe samples of galaxies where we analyzed the possible accretion disk contribution to the total flux of their broad emission lines. We review the methods of analyzes in Sections 3 and in Section 4 we give results. Finally, in Section 6 we outline our conclusions.

\section{Data sets}

First sample of 12 AGN spectra was observed  with the 2.5 m INT on La Palma at
2002 \citep{Pop04}. 
For some lines with bad S/N ratio we used HST observations obtained with the Space Telescope Imaging Spectrograph (STIS) on January 2000 (NGC 3516). Details about observation and data reduction can be find in \citep{Bon06,Pop04}
The narrow and satellite lines were cleaned \citep[see e.g.][]{Pop02,Pop03,Pop04}.

With the idea to compare our model with real emition line profiles, without fitting, we used the { sample of spectra containing }90 broad-line-emitting AGN, which have been collected by \citet{Gio07} from the third data release of Sloan Digital Sky
Survey (SDSS)\footnote{http://www.sdss.org/dr3}.
The spectra were already corrected for sky-emission, telluric absorption, the Galactic extinction and redshift \citep{Gio07}.
Since the interest was to investigate the broad emission line profiles, we
subtracted the narrow components of H$\alpha$ and the
satellite [NII] lines.
The spectral reduction (including subtraction of stellar component) and the way to obtain
the  broad line profiles are in more details explained in \citet{Gio07}.

Previously cleaned broad H$\alpha$ profile was
normalized, converted from wavelength to velocity scale, using DIPSO software package\footnote{http://www.starlink.rl.ac.uk}.

{ The sample of 90 AGN contained the spectra with different FWHM of broad H$_{\alpha}$ from about 1000 km/s to 7000 km/s, with at least several representatives in every 500 km/s \citet[see Table 2 in][]{Gio07}. This covered many different types of broad single-peaked profiles.}

\section{Methods of BEL profile analyzes}

We performed several tests to find a possible  accretion disk contribution to the total line flux of a single peak broad emission lines. \\

\subsection{Gaussian fit} 
 We used a $\chi^2$ minimization routine to obtain the best fit. 
To limit the  number of free parameters in the fit, we have set some
{\it a priori}  constraints as given in \citet{Pop03}. 

\begin{figure}
  \centering
  \includegraphics[angle=0,width=7cm]{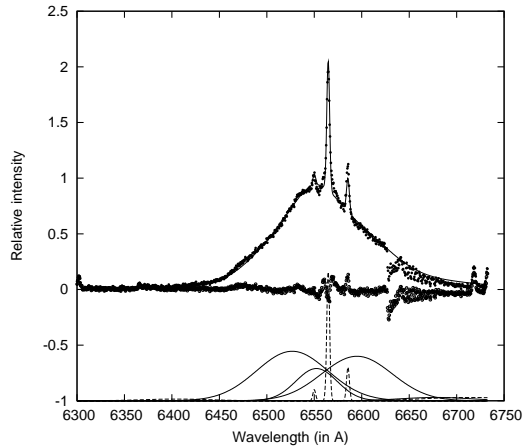}

\caption{Decomposition of the
H$\alpha$  line of Mrk 841.  The dashed lines represent
the observations, while the solid lines show the profile
obtained by Gaussian decomposition.  The Gaussian components are
presented at the bottom. The narrow dashed lines correspond to the narrow H$\alpha$ and
[NII] lines.}
\label{fit-gauss-Ha}
  \end{figure}

\begin{figure}
  \centering
  \includegraphics[angle=270,width=7cm]{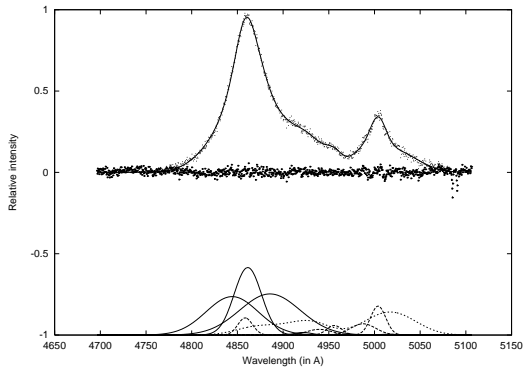}

\caption{Decomposition of the
H$\beta$  line of 3c 273.  The dashed lines represent
the observations while the solid lines show the profile
obtained by Gaussian decomposition.  The Gaussian components are
presented at the bottom. The narrow dashed lines correspond to the narrow H$\beta$,
[OIII] lines and FeII shelf.}
\label{fit-gauss-Hb}
  \end{figure}

{ In the fitting procedure, we looked for the minimal number of Gaussian components
needed to fit { the broad component}. { For the narrow blending lines we have been} taking into account the intensity ratio of [OIII] lines ({ obtained form} the atomic values 1:3.03) and the fit of Fe II template for the H$_\beta$ In the case of H$_\alpha$ line, for the narrow [NII] lines we assume that their intensity ratio is 1:2.96 \citep[see][]{Pop03,Pop04,Bon06}. Some examples are presented in Figures \ref{fit-gauss-Ha} and \ref{fit-gauss-Hb}.
}

The results of indicated existence of two kinematic regions, one very broad line region (VBLR), represented by the two shifted Gaussians and one intermediate broad line region (ILR), represented with the central Gaussian component. It was noticed that both shifted components had similar widths, while the central component had similar values to the central components of the other lines (see Fig. 3), what led to idea to introduce the two component model, that consists of the accretion disk (which could describe the wings of the line), and surrounding region with isotropic velocities of clouds (representing the core of the line, that can be described by a Gaussian).

\begin{figure}
  \centering
  \includegraphics[angle=0,width=7cm]{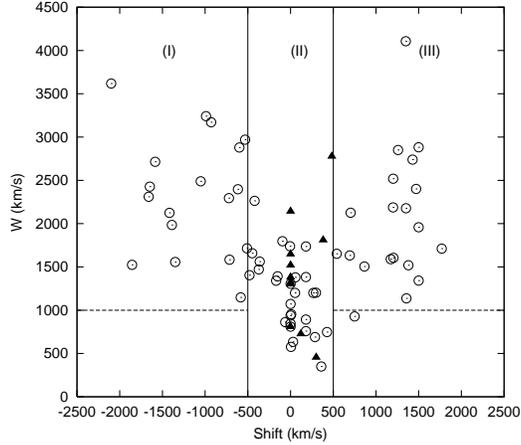}
     \caption{The widths (w) as a function of the shifts of the broad
Gaussian components obtained for the H$\alpha$ and H$\beta$ lines
of our sample of AGNs (open circles) and the Fe II template (full
triangles) \citep[][]{Pop04}.}
\label{widths}
\end{figure}

These results encouraged us to test in the same manner the sample of 12 AGN \citep{Pop04}. After Gaussian decomposition the results showed again two distinct kinematic regions. 
The results of Gaussian analysis showed that Fe II lines had comparable widths to the central component (see Fig. \ref{widths}), indicating that they originate in the same region, { what was previously shown by \citet{Marziani03} and recently by \citet{Hu08}.}

\subsection{Fit with a two-component model}

For the disk emission, we adopted disk model developed by \citet{Che1}. We fitted the profiles with composition of the disk and Gaussian profiles (see Figures 4-6). Starting value of the width of the Gaussian was considered to be the value of the width of the central Gaussian obtained from the fit with three broad Gaussians components, for the same object. All those lines were expected to have a common inclination, since they originate in the same object, so this constrain was taken into account.\\

Using these results as a starting point, we fitted with the two-component model the profile composed as averaged H$\alpha$ and H$\beta$ line of each AGN \citep[see][]{Pop04}.

The results of the fit are presented in Figs 4-6, and 
corresponding parameters of two component model are given in Tables 1-3 for 12 AGN.
As one can see from Figs 4-6, the line profiles can be well fitted
with a two-component model. Depending on the inclination, local broadening
and the dimension of the accretion disk, a line from the disk can appear
as single or double-peaked. 

The parameters obtained by fitting the line profiles are given in the
Tables 1-3 for 12 AGN. Here we presented several tests of fitting with two component model. We fixed the  emissivity parameter as p=3 (Table 1) and also  as p=2.5 (Table 2), with allowing all other parameters to be free. While searching for the maximal inclination of the disk, we obtained values of parameters presented in Table 3. { For the parameter that corespond to the shift of the disk component, { there are several possible explanations, like a motion of emitters in elliptical disks \citep{Era95}, tidal perturbation of the disk around a supermassive black hole by a smaller companion \citep{Era95}, 
or some binary black hole configurations in which BLR is dominated by the gravity of one black hole. The orbital motion of two black holes around their center of mass can result in radial velocity shift of the entire BLR  \citep[see for example][]{Boroson09,Bogd09}. } Moreover, one can expect different types of turbulence caused by magnetic field in the accretion disk \citep[see e.g.][]{Turner2003} that in the case of the line additionally contribute to the line broadening. The magnetic field could produce the effects of the accretion inflow and outflow around the back hole \citep[see. e.g.][]{Ohsuga2009}, that also can partly contribute to the width and shift of a line originated from a small disk cell.}

   \begin{figure}
   \centering
   \includegraphics[angle=0,width=7cm]{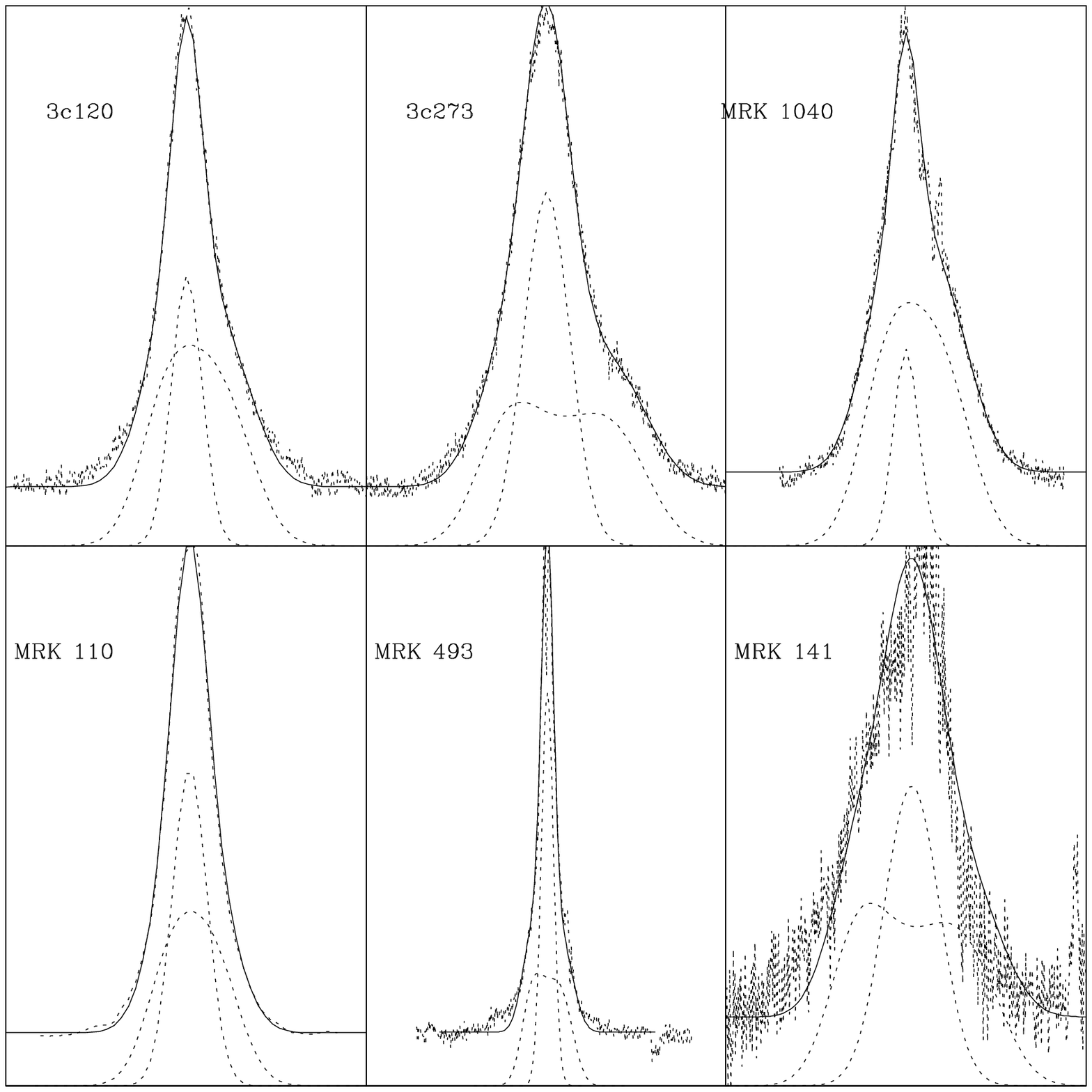}
  \includegraphics[angle=0,width=7cm]{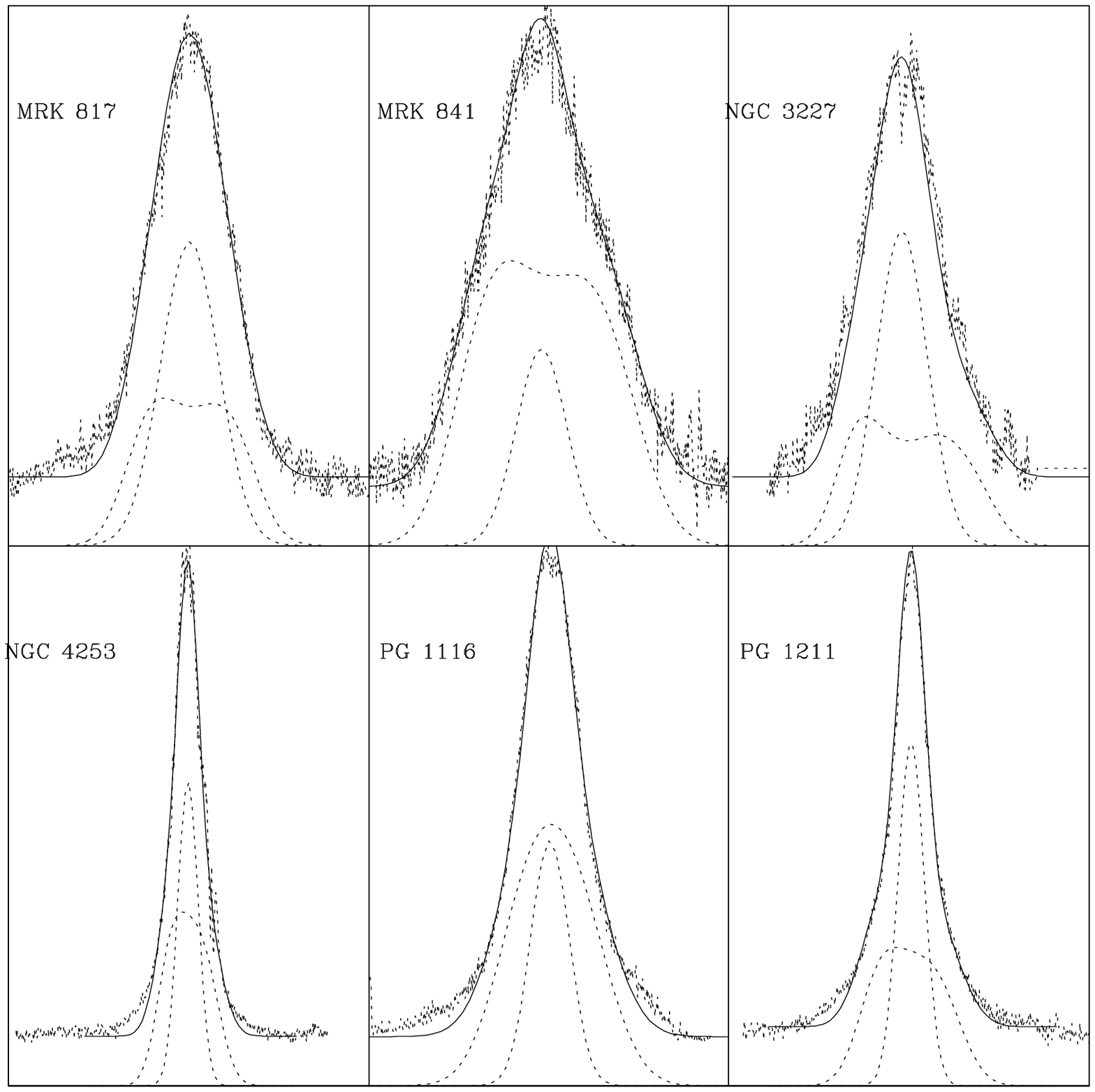}
      \caption{Observed averaged H$\beta$ and H$\alpha$ lines
 of {12} AGNs (dashed lines),  fitted with the two-component model (solid
lines) which
include the
emission of an accretion disk and a spherical emission region as a broader
or double-peaked contour or a narrower Gaussian, respectively (dashed
lines at bottom).
              }
         \label{FigVibStab4}
   \end{figure}

Concerning the disk parameters we can  point out the following:
  (i) The outer radii are very
similar in the sample (R$_{\rm out}\sim 1500\ R_g$);
(ii)  The local random velocities in the disk are different from object to
object and they are in the interval from 400 km/s to 1300 km/s.
(iii) The inner radii of the emitting disk for all considered objects are in the interval from
300 to 600 R$_g$,
(iv) The obtained inclinations are small ($5^\circ<i<15^\circ$), and such
values of inclinations support the idea  that we are in
position to more frequently observe the Sy 1 at   face-on
inclinations.

\begin{table*} 
\begin{center}
\caption{The parameters of the disk: z$_{\rm disk}$ is the 
shift and $\sigma$ is the  Gaussian broadening term
from disk indicating the random velocity in disk, R$_{\rm inn}$ are the
inner radii, R$_{\rm out}$ are the outer radii. The z$_{\rm G}$ and W$_{\rm
G}$ represent the parameters of the Gaussian component.  $F_s/F_d$ represents the ratio of
the relative disk and Gaussian fluxes. The value for parameter of emissivity was fixed as p=3.}
 
\begin{tabular}{|c|c|c|c|c|c|c|c|c|c|}
\hline

 Object &$i$& z$_{\rm disk}$  &$\sigma$ (km/s) & R$_{\rm inn}\ (R_g)$ 
& R$_{\rm out}\ (R_g)$  & z$_{\rm G}$ & W$_{\rm G}$ (km/s)& $F_s/F_d$\\
\hline

 3C 120 & 10 & -300 & 1060 &  380 & 1400 & +30 & 900 & 0.6  \\
 3C 273 & 14 & -30 & 1140 & 400  & 1400 & +30  & 1350& 0.8 \\
MRK 1040&  8& -250  & 1050 & 400& 1400 & 00 & 750&  0.2 \\
MRK 110&7 & -320 & 950 & 500 & 1400  & +150 & 1020& 0.9 \\
MRK 141 &12 & -630 & 1050 & 300 & 1100  & +200  & 1620& 0.7 \\
MRK 493 &5 &  -480 & 400 & 600 & 1400  & +60  & 360& 1.0 \\
MRK 817 &12 & -450 & 1050 & 600 & 1400  & +48  & 1600& 0.95 \\
MRK 841 &15 &-750 &  1270 & 450 & 1400  & -300  & 1500& 0.2 \\
NGC 3227&12 & -780 & 1080 & 350 & 1600  & +300  & 1400& 0.8 \\
NGC 4253&5 & -630 & 560 & 500 & 1500  & -30  & 600& 0.7 \\
PG 1116 &8 &  -450 & 1260 & 500 & 1400  & +48  & 1140& 0.4 \\
PG 1211 &8 & -660 & 830 & 430 & 1530  & +90  & 780& 0.8 \\
\hline

\end{tabular}
\end{center}
\end{table*}

\begin{table*} 
\begin{center}
\caption{The same as in Table 1, but with fixed value for parameter of emissivity as p=2.5}
\begin{tabular}{|c|c|c|c|c|c|c|c|c|}
\hline
Object & $i$ & z$_{\rm disk}$  & $\sigma$ (km/s)   &R$_{\rm inn}$ ($R_g$)& R$_{\rm out}$ ($R_g$) & z$_{\rm G}$ & W$_{\rm G}$ (km/s)& $F_s/F_d$\\
\hline
3C 120 & 14  & +300            & 960               &   650               &     2500              &   +300      &      960          &      0.7 \\
3C 273 & 18  & +90             & 1191              &   550               &     1900              &   +60       &     1380          &      0.9 \\
MRK1040& 12  & 0               & 870               &   700               &     2700              &    00       &      600          &      0.2 \\
MRK 110& 12  & 0               & 890               &   800               &     2400              &   +150      &     1020          &      1.4 \\
MRK 141& 14  & -630            & 1040              &   300               &     2200              &   +240      &     1620          &      0.8 \\
MRK 493&  5  & -300            & 280               &   800               &     2400              &   +60       &     330           &      0.9 \\
MRK 817& 14  & -300            & 1042              &   600               &     2400              &   +48       &     1690          &      1.2 \\
MRK 841& 23  & -300            & 1270              &   950               &     2400              &   -300      &     1440          &      0.3 \\
NGC3227& 19  & -360            & 890               &   800               &     2600              &   -300      &     1500          &      1.0 \\
NGC4253& 8   & -150            & 595               &   1500              &     9500              &    -90      &     510           &      0.4 \\
PG 1116&15   &  0              & 1480              &   850               &     3800              &     +0      &     1440          &      1.0 \\
PG 1211&12   & -240            & 680               &   730               &     2530              &    +90      &     780           &      0.9 \\
\hline
\end{tabular}
\end{center}
\end{table*}


\begin{table*} 
\begin{center}
\caption{The same as in Table 1, but in search for maximal inclination}

\begin{tabular}{|c|c|c|c|c|c|c|c|c|c|}
\hline
Object & $i$ & z$_{\rm disk}$  & $\sigma$ (km/s)   &R$_{\rm inn}$ ($R_g$)& R$_{\rm out}$ ($R_g$) & z$_{\rm G}$ & W$_{\rm G}$ (km/s)&  p &$F_s/F_d$\\
\hline
 3C 120 & 30 & +360 & 745 &  1650 & 19000 & +60 & 960 &2.2 & 0.5\\
 3C 273 & 30 & +330 & 490 & 1250  & 15000 & +60  & 1380& 2.8 & 0.9\\
MRK 1040&  27& +300  & 700 & 100& 17200 & 00 & 480& 1.3 & 0.24 \\
MRK 110&30 & +210 & 280 & 1800 & 22400  & +150 & 900& 2.0  & 1.2\\
MRK 141 &33 & -450 & 530 & 1200 & 1000  & +300  & 1620& 2.1 & 0.8\\
MRK 493 &30 &  +60 & 255 & 10800 & 124000  & +60  & 270& 1.8 & 0.7\\
MRK 817 &35 & 0 & 600 & 800 & 14400  & +50  & 1500& 1.9 & 1.0\\
MRK 841 &50 &-150 &  765 & 1950 & 27400  & -300  & 1440& 2.1 & 0.8\\
NGC 3227&34 & -300 & 620 & 1300 & 12600  & -300  & 1500& 2.1 & 0.77\\
NGC 4253&25 & -90 & 215 & 2500 & 69500  & -30  & 510& 2.0 & 0.6\\
PG 1116 &30 &  0 & 850 & 950 & 15800  & +90  & 1440& 2.2 & 0.3\\
PG 1211 &30 & 0 & 383 & 920 & 15530  & +90  & 780& 1.9 & 0.2\\ 
\hline
\end{tabular}
\end{center}
\end{table*}

Concerning the emission of the surrounding region we can point out
that: i) the red-shifts are close to the cosmological,  they
are in the interval of $\pm$ 300 km/s; (ii) the random velocities in this region are
also different for different objects, and they are in the interval
from 400 to 1600 km/s.

There are indications that local broadening in the disk and in surrounding region could be correlated (see Tables 1-3). It could indicate that those parameters could be linked by the same process, like wind-disk interaction, or something similar.

On the other side, we compared the disk parameters obtained from the two-component model with the
ones obtained by fitting the double-peaked line profiles with the disk model only. \citet{Era} fitted  a  sample of
12  double peaked AGNs with the disk model. 
In comparison with their results, our disk parameters are with smaller inclination and also
smaller outer radii. \citet{Str03} investigated the spectra of 116 AGNs with double peaked lines using the disk model. 
and obtained inclinations of the accretion disks smaller than 50$^\circ$,  while inner radii were from 200 R$_g$ to 800 R$_g$ and  local turbulent velocities were from 780 km/s  to 1800 km/s, that well fitted our estimated values (see Tables 1-3). Comparing those parameters with ones in Tables 1-3, we can conclude that { they} could be in a good agreement, even though  {  our inclinations were significantly smaller than
upper value of those given by the authors}. Moreover, for double-peaked
lines one can expect a higher inclination, { because the
value of inclination strongly affects the line width, even for smaller values of emissivity index}. 
For the outer radii, the estimated values were larger
than 2000 R$_g$ \citep{Era, Str03}. { Our results showed}  smaller { values of} outer radii in comparison with { those} estimated from \citep{Era, Str03}, (see Tables 1-3).

\subsection{Estimation of the  accretion disk contribution to the BELs}

Due to a large number of free parameters, we were not able to determine an unique solution  of parameters for the best fit (see Figs 3-6). Therefore, we introduced alternative approach to the analysis of the disk contribution to the single peak broad emission lines.

First we set some constrains in the parameter space, and we constructed {a} grid of simulated line profiles. As a rough approximation, we fixed the width of the central Gaussian component to the 1000 km/s, as mean value obtained from the two component model fit \citep[][]{Bon06}. As it was shown in \citet[][]{Pop04}, the random velocities in the disk and in the non-disk region were approximately the same, so it was also fixed to the value of 1000 km/s.

\begin{figure}
   \centering  
   \includegraphics[angle=0,width=7 cm]{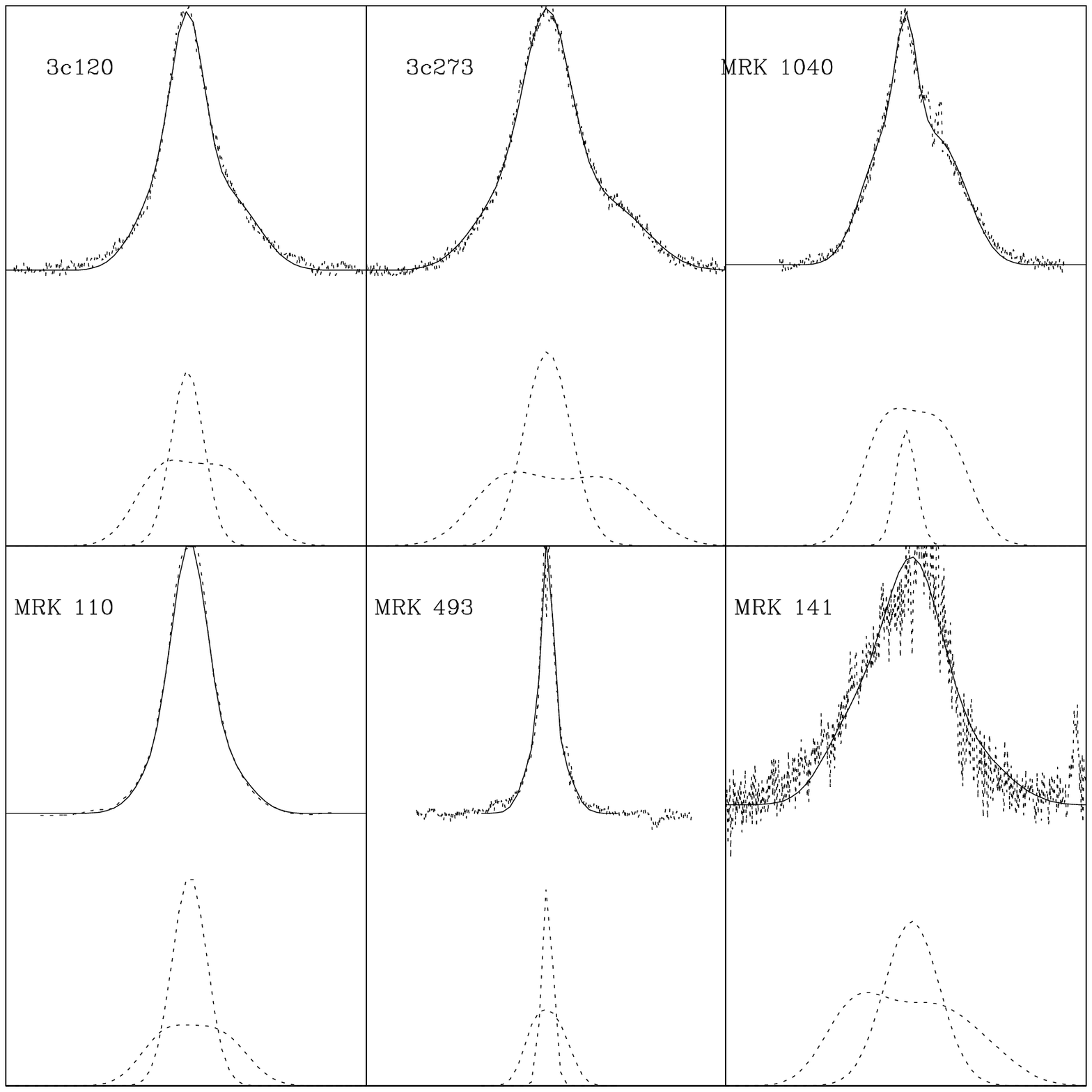}
   \includegraphics[angle=0,width=7 cm]{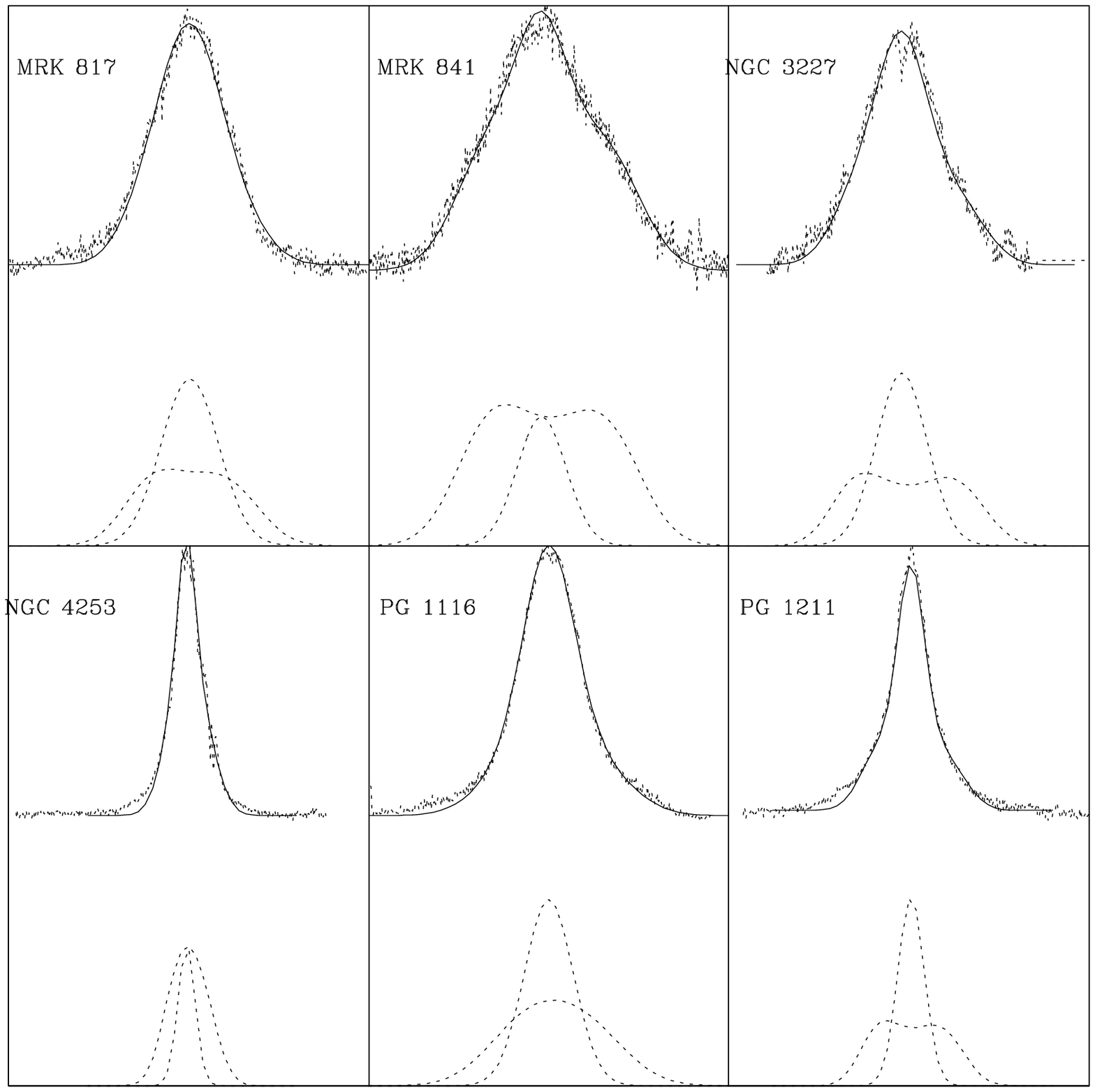}
      \caption{The same as in Fig. { 4}, but for p=2.5.
              }  
         \label{FigVibStab}
   \end{figure}

\begin{figure}
   \centering
   \includegraphics[angle=0,width=7 cm]{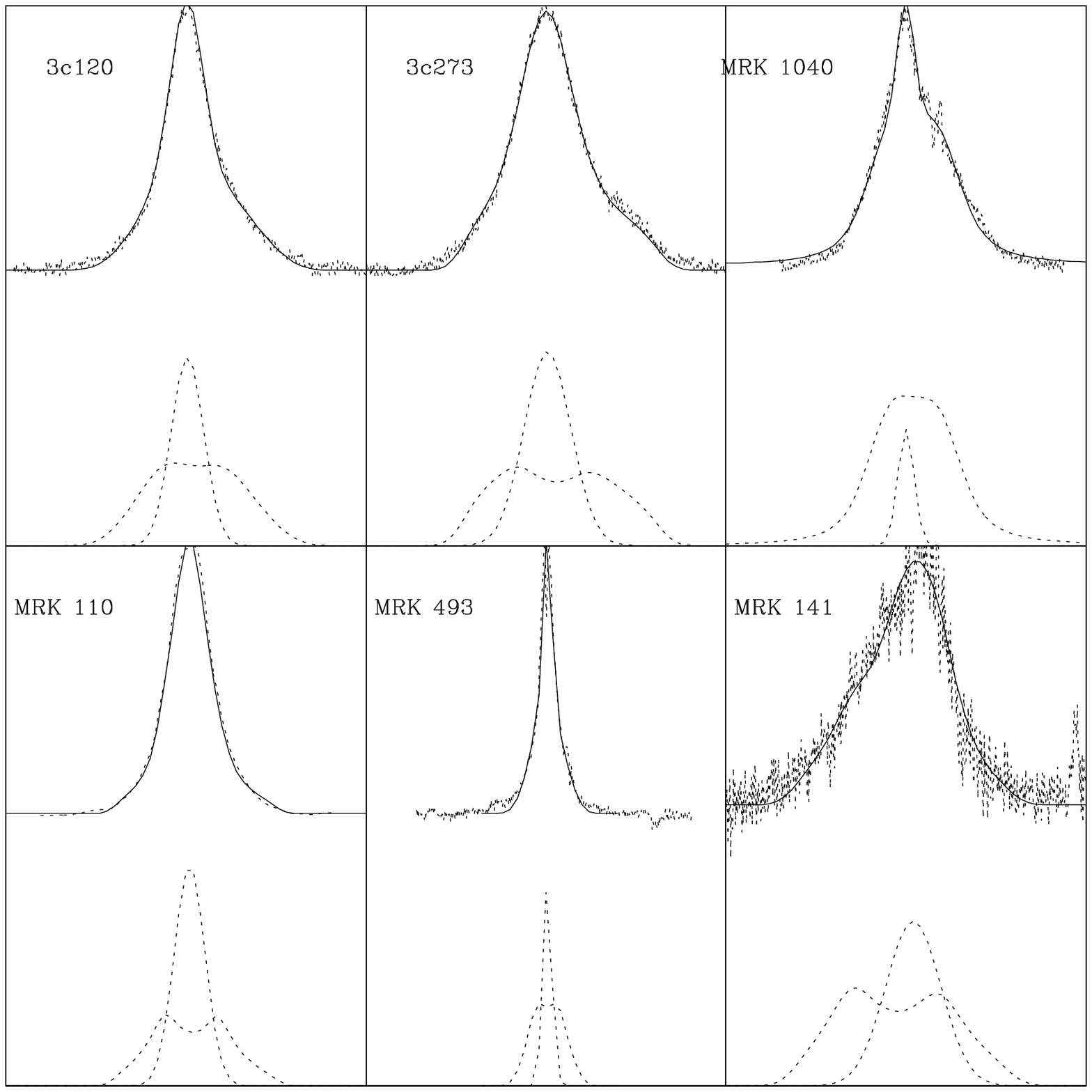}
   \includegraphics[angle=0,width=7 cm]{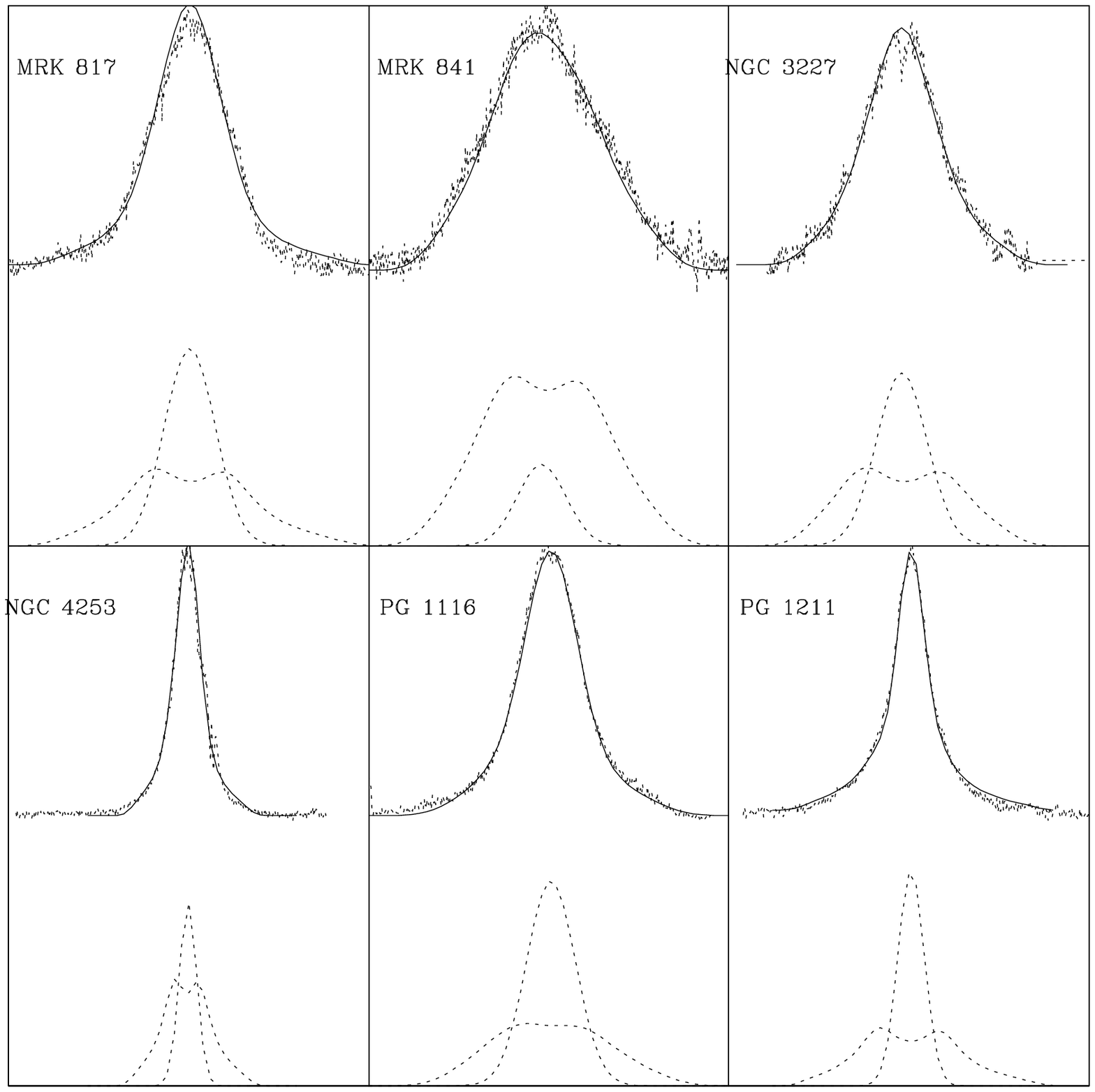}
      \caption{The same as in Fig. { 4}, but for maximal value of
inclination, and with constraints that $i>25$.
              }
         \label{x}
   \end{figure}

The flux ratio of the simulated profiles was given by the parameter $Q={F_{\rm{s}}\over F_{\rm{disk}}}$, where $F_{\rm{s}}$ is the flux emission from the spherical region and $F_{\rm{disk}}$ is the flux emission from the accretion disk. In the simulations
the different values of the parameter $Q$ have
been considered, with the composite profile normalized to unity. The emissivity of the disk as a function of radius, $r$, is given by $\epsilon=\epsilon_0r^{-p}.$

We created the grid of simulated BEL profiles with different parameters of the two component model. To reduce free parameters we estimated some constrains. Due to small change in profiles we could fix the following parameters:

\begin{itemize}
\item the emissivity of the disk was fixed to p=3,
\item the Gaussian broadening in the disk was fixed to 1000 km/s,
\item the random velocities of the isotropic component was also fixed to 1000 km/s,
\item there were no big differences in shape of the profiles if the outer radius was higher then several thousands R$_g$,
\item as {a mean value of results obtained} from the fit,  we fixed the size of inner and outer radius for the first sample of 12 AGN R$_{inn}$=400 R$_g$ and R$_{out}$=3000 R$_g$, while for the second sample of 90 SDSS AGN, these parameters were fixed by averaging the disk size obtained from the fitting of the double-peaked lines in the work of \citet{Era1}, so we fixed  the inner radius to be R$_{inn}$=600 R$_g$, and the
outer R$_{out}$=4000 R$_g$.
\end{itemize}

It was noticed that for the change of the shape of BEL profiles, variation of inclination and the flux ratio played dominant role,  so the grid of models was finally reduced only to these two parameters. Simulated spectra were compared  to the spectra from different samples.

First the profiles were
 normalized to the same intensity. {We  noticed that the variation of the emissivity parameter  could 
be neglected since normalized profiles  showed very 
small change in shape} \citep[see][]{Bon09,Bon08}. For that reason, this parameter was fixed to the value p=3.

For each line profile,  the full widths at 10\%, 20\%, 30\% of maximum
intensity were measured, and compared after being normalized to
full width at half maximum, which was introduced through coefficients
$k_j$ ($j=10,\ 20,\ 30$) as $k_{10}=w_{10\%}/w_{50\%},\
k_{20}=w_{20\%}/w_{50\%}$ and $k_{30}=w_{30\%}/w_{50\%}$.

Fixing the inner and outer radius to an averaged value, obtained from the study
of BELs with double-peaked lines \citep{Era1}, we estimated the values of $i$ and
$Q$ for all analyzed galaxies.

\begin{figure}
\centering
\includegraphics[width=8.5 cm, angle=270]{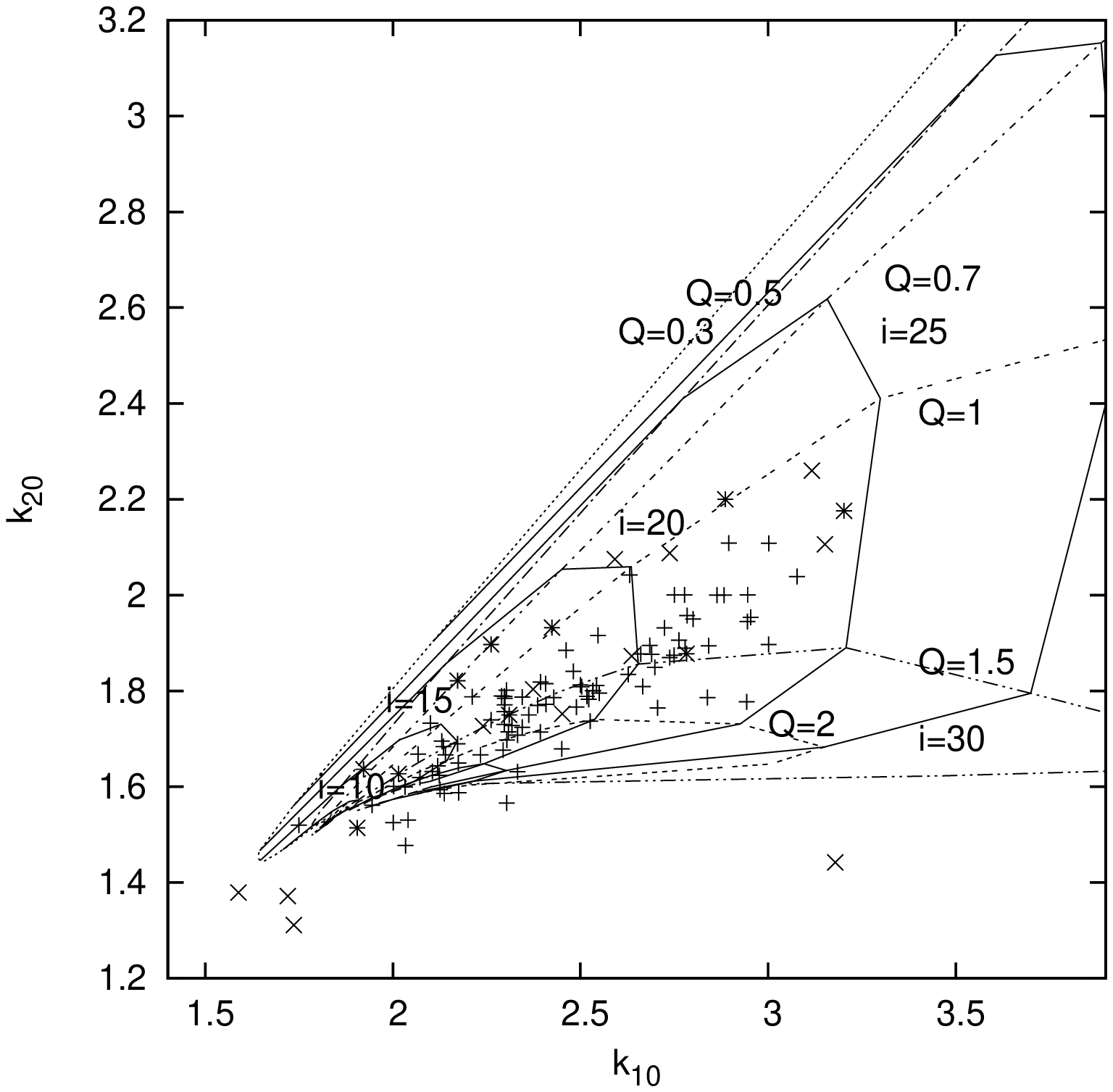}
\includegraphics[width=8.5 cm, angle=270]{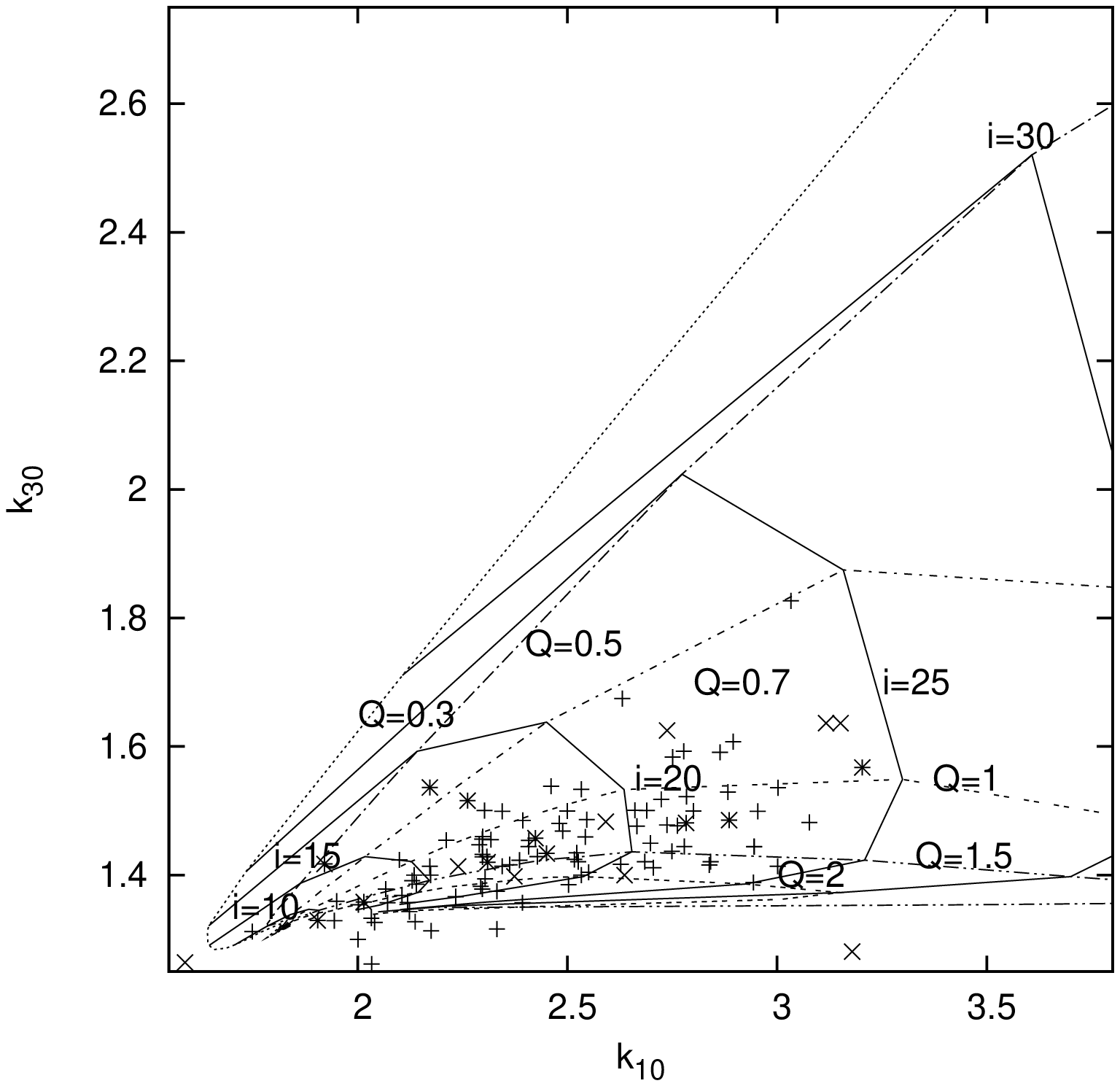}
\caption{The measured width ratios  (crosses-sample of 90 SDSS, asterisks-sample of 12 H$_{\alpha}$, x signs-sample of 12 H$_{\beta}$) and simulated values (dashed lines) from the two-component model for the different  
contribution of the disk emission to the total line flux ($Q$=0.3, 0.5, 0.7, 1, 1.5 { and 2}) and different
inclinations are considered (solid  iso lines  presented   {{$i$ = 10, 15, 20, 25 and 30}} degrees, respectively).}
\label{merenja}
\end{figure}

For the first sample, the inner radius of the disk was chosen to be $R_{inn}=400\ R_G$, {as the mean value
of this parameter in the previous fittings} \citep{Bon08}.
For the outer radius, the chosen value was $R_{out}=3000\ R_G$, although the {mean value of the}  outer radius  for the sample was about 30000 \citep{Bon06}, {since the} variation of this parameter, after {values of} several thousand gravitational radii, {did not show} significant influence {to} the shape of the line (especially if the line is noisy).

\section{Results}

In order to estimate possible disk presence in the single peaked broad emission lines of AGN, we applied several methods.

From the Gaussian analysis we could conclude that two distinct kinematic regions might be present in the BELs. The central Gaussian component indicated the existence of the emission from the region with intermediate velocities (ILR), while in the wings two shifted broad Gaussians indicated possible presence of disk-like emission (VBLR). The results for the widths of the emission lines of Fe II template showed that these lines probably originate in the ILR (see Figures 1-3).

From the fit with two component model where one component represents the emission from the accretion disk (mostly affects the wings of the emission line), and another from the surrounding region with isotropic velocity distribution of the emitters (that contribute to the core of the line), we can conclude:

\begin{itemize}
\item two component model can well fit the profiles of broad emission lines {of our samples}, but it is very hard to determine the disk parameters, because of the large number of the free parameters (see Figs 4-6),
\item random velocities in the surrounding region with isotropic velocities are similar to random velocities in the disk, what implies  that these two regions could be connected through some process (for example through the wind of the accretion disk),
\item the values of the parameters indicate that the inclinations are smaller than 50$^\circ$, and the dimensions of the disk could be from several hundred $R_g$ for the inner radius, to several hundred thousands $R_g$ for the outer radius.

\end{itemize}

After comparing the measurements of $k_i$ for the sample, with the measurements of the simulated profiles, we could derive the estimates of $i$ and $Q$. As the result, we found that the most of the measured points were
between $0.5<Q<1.5$, and $10^\circ<i<25^\circ$ (see Fig. \ref{merenja}). The results showed a consistence between estimations of $i$ and $Q$ for both H$\alpha$ and H$\beta$, with very small discrepancies
\citep[see Table 1 in][]{Bon08}. 

After this analysis, we tested this method to the larger sample of 90 AGN \citep[see][]{Bon09}. The measurements were performed in the same way as with 12 AGN, with the difference that we compared it to the simulated spectra with
the disk dimensions of $R_{inn}=600$ Rg and ${R_{out}=4000} \ {\rm Rg}$ (averaging the disk size obtained from the fitting of the double-peaked lines in the work of \citep{Era1}). With this sample we measured only H$\alpha$ lines.
As it could be seen  in Fig. \ref{merenja}, most of the measurements are located within $1<Q<2$, and $10^\circ<i<25^\circ$  \citep[see][]{Bon09}.

\section{Conclusions}

In this paper we outline our recent investigation about the presence of the disk emission in single peaked BELs.

We {carried out the} following analyzes:
i) fitting of single peaked BELs with Gaussians and with two component model, and
ii) comparing line parameters of a sample with simulated profiles.

From our investigations we can conclude that there could be a high probability that the disk emission flux might be present in the single peaked emission line profiles. In principle, the contribution of the disk emission to the total flux of the single peaked BELs is in most cases smaller then 50\%. Also, we found that the disk inclinations were mainly smaller then  $i<25^\circ$. Such small inclinations and large discrepancies from simulated profiles for higher inclinations should be discussed in the contest of unified model, i.e. possible disk orientation to the torus or partial obscuration by the torus.

\section*{Acknowledgments}

The work was supported by the Ministry of Science and Technology
of Serbia through the project 146002: ``Astrophysical spectroscopy of extragalactic objects``.
We would like to thank to Jack W. Sulentic, Paola Marziani, Michael Eracleous and Martin Gaskell for many helpful discussions and comments.




\end{document}